# Processor in Non-Volatile Memory (PiNVSM): Towards to Data-centric Computing in Decentralized Environment

Viacheslav Dubeyko

*Abstract*— The AI problem has no solution in the environment of existing hardware stack and OS architecture. CPU-centric model of computation has a huge number of drawbacks that originate from memory hierarchy and obsolete architecture of the computing core. The concept of mixing memory and logic has been around since 1960s. However, the concept of Processor-In-Memory (PIM) is unable to resolve the critical issues of the CPU-centric computing model because of inevitable replication of von Neumann architecture's drawbacks. The next generation of NVM/SCM memory is able to give the second birth to the data-centric computing paradigm. This paper presents a concept of Processor in Non-Volatile Memory (PiNVSM) architecture. The basis of PiNVSM architecture is the concept of DPU that contains the NVM memory and dedicated PU. All necessary PU's registers can be implemented in the space of NVM memory. NVM memory of DPU is the single space for storing and transformation of data. In the basis of PiNVSM architecture lies the DPU array is able to overcome the limitations as Turing machine model as von Neumann architecture. The DPU array hasn't a centralized computing core. Every data portion has dedicated computing core that excludes the necessity to transfer data to the place of data processing. Every DPU contains data portion that is associated with the set of keywords. Any complex data structure can be split on elementary items that can be stored into independent DPU with dedicated computing core(s). One DPU is able to apply the elementary transformation on one item. But the DPU array is able to make the transformation of complex structure by means of concurrent execution of elementary transformations in different DPUs. The PiNVSM architecture suggests a principally new architecture of the computing core that creates a new opportunity for data self-organization, data and code synthesis.

## I. INTRODUCTION

The next generation of NVM/SCM memory would be able to open the new horizons for future computer technologies. Persistent byte-addressable memory of huge capacity with low latency and good endurance is the holy grail of computer science that is able to change the game dramatically. However, currently, the NVM/SCM memory represents the big challenge but not the hope to resolve the computer science's problems. It is well known fact that the amount of data is growing exponentially. Every day makes the problem of data processing more and more crucial. But the architecture of modern CPUs and programming model of data processing are the critical bottlenecks for the case of Big Data. The AI problem has no solution in the environment of existing hardware stack and OS architecture. Even if anybody imagines that AI problem is solved theoretically then it will need to use the resources of several datacenters for the real implementation of AI instance in the current data processing paradigm. But, most probably, even several datacenters will have not enough resources for the implementation of AI instance. The power consumption of such implementation of AI instance would be enormously huge. It is possible to point out that current state of the art is unable to provide the basic hardware technologies for implementation of AI instance.

This work presents a concept of Processor in Non-Volatile Memory (PiNVSM) architecture. The basis of PiNVSM architecture is the concept of DPU that contains the NVM memory and dedicated PU. All necessary PU's registers can be implemented in the space of NVM memory that can decrease the power-consumption dramatically. NVM memory of DPU is the single space for storing and transformation of data. In the basis of PiNVSM architecture lies the DPU array is able to overcome the limitations as Turing machine model as von Neumann architecture. The DPU array hasn't a centralized computing core. Every data portion has dedicated computing core that excludes the necessity to transfer data to the place of data processing. The basis of PiNVSM paradigm is not algorithm but data itself. Every DPU contains data portion that is associated with the set of keywords. Generally speaking, DPU array creates deeply decentralized paradigm of data processing. Any complex data structure can be split on elementary items that can be stored into independent DPU with dedicated computing core(s). Finally, it means that one DPU is able to apply the elementary transformation on one item. But the DPU array is able to make the transformation of complex structure by means of concurrent execution of elementary transformations in different DPUs. The PiNVSM architecture suggests a principally new architecture of the computing core that creates a new opportunity for data self-organization, data and code synthesis.

This paper makes the following contributions:

- It was introduced the concept of Processor in Non-Volatile Memory (PiNVSM) architecture. The whole system could be represented by DPU arrays. Every DPU contains the NVM memory and dedicated PU. NVM memory of DPU is the single space for storing and transformation of data. Every DPU contains data portion that is associated with the set of keywords. Any complex data structure can be split on elementary items that can be stored in independent DPU with dedicated computing core(s). One DPU is able to apply the elementary transformation on one item. But the DPU array is able to make the transformation of complex structure by means of concurrent execution of elementary transformations in different DPUs.
- It was discussed how the PiNVSM architecture is able to resolve the issues of CPU-centric computing paradigm.

The rest of this paper is organized as follows. Section II presents the background and motivation of the paper. Section III surveys the related works. Section IV introduces the concept of Processor in Non-Volatile Memory (PiNVSM) architecture. Section V contains the discussion and analysis of potential advantages of the PiNVSM architecture. Section VI offers conclusions.

## II. BACKGROUND & MOTIVATION

### A. Current Stack Issues for the Case of NVM/SCM Memory

The next generation of NVM/SCM memory would be able to open the new horizons for future computer technologies. Persistent byte-addressable memory of huge capacity with low latency and good endurance is the holy grail of computer science that is able to change the game dramatically. However, currently, the NVM/SCM memory represents the big challenge but not the hope to resolve the computer science's problems. What problems do the next generation of NVM/SCM memory emerge for the computer scientists and practitioners? First of all, it needs to take into account the current state of the art. The CPU core transforms data by means of small set of really fast registers. This set of registers was inherited from that times when the whole capacity of memory in the system was comparable with number of registers in the CPU core. Also it is worth to point out that all modern programming languages are based on the programming model is taking into account the limited set of CPU core's registers. As a result, the key assembler instruction is the exchanging by data between the main memory and registers of CPU core. This obsolete model of limited set of CPU core's registers was enhanced by adding the L1/L2/L3 caches into the CPU core with the goal to improve the system performance. These L1/L2/L3 caches provide the way to keep the frequently accessed data and predicted or expected instructions sequence near the CPU core. However, these caches create the problem of coherence of data between the main memory and the caches of different CPU cores. Generally speaking, the modern CPU spends a lot of time for data/code exchange between the main memory and CPU cores. The next important problem is the main memory. At the time of invention the first computer technologies the persistent media was very slow and the DRAM memory was so expensive that it was elaborated the key compromise of creation the memory hierarchy. This compromise implies that small portion of accessed or requested data is stored temporary in the fast volatile memory but the main snapshot of data state is kept into the slow persistent storage. It means that the fast volatile memory plays the role of window or mediator between the CPU core and slow persistent storage. Generally speaking, this principal dichotomy between volatile and persistent memory has fundamental nature in the modern OS and it cannot be resolved by simple addition of a new OS's subsystem. File system plays the principal role in the OS by means of copying or pre-fetching data from persistent storage into the main memory and synchronization of modified data between volatile memory and the file system volume on persistent media. The tricky problem of modern OS is that any file system provides very important features (namespace, permissions) via the inevitable necessity to copy and convert metadata between persistent storage and internal OS's metadata representation in the volatile memory. Moreover, any user-space application initiates a huge amount of metadata operations by means of requesting the metadata from the file system volume and updating the metadata on persistent media. As a result, simple exchange the volatile memory (DRAM) on non-volatile memory cannot resolve the problem because the internal mechanisms of OS and file system require the presence of two levels of memory (as minimum) and exchanging by data between of them.

### B. Big Data Challenge

Big Data is a big challenge. It is well known fact that the amount of data is growing exponentially. Every day makes the problem of data processing more and more critical. First of all, the problem of Big Data is the problem of storing such immense volume of data. The next, and maybe more critical, is the problem of data processing for reasonable time with the representative results. The first critical point is the storing as much as possible data inside of one local node or local storage device. Otherwise, the possible critical bottleneck would be the necessity in intensive data exchange via network between remote nodes. However, the one node's resources are limited and even the distributed data processing in the network of many nodes can be limited by impossibility to exceed some threshold of available resources. It means that some Big Data problems cannot be processed at all in the environment of current state of the art. It is possible to suggest to use a huge shared pool of memory. However, this approach still is unable to resolve the problem of intensive exchange by data between the shared memory pool and CPU caches. Also the problem of data coherence becomes more severe for the case of the shared memory pool. Another very critical problem would be the problem of data addressing in the huge shared memory pool. Because, first of all, any CPU is able to address the limited virtual memory space. Secondly, the latency of different addresses in the huge memory pool would be significantly heterogeneous. Finally, the performance of an application could vary significantly for different ranges of addresses. But, do we really need in the huge shared memory pool? The problem of storing as much as possible data for a media's area unit is very important. But the increasing this number has very unpleasant side effect of I/O performance degradation in the current paradigm of I/O stack. Generally speaking, such degradation takes place because of necessity to exchange data between CPU core and persistent storage device by means of the intermediate main memory (DRAM). As a result, the architecture of modern CPUs and programming model of data processing are the critical bottlenecks for the case of Big Data.

## C. AI Challenge

The AI problem has no solution in the environment of existing hardware stack and OS architecture. Even if anybody imagines that AI problem is solved theoretically then it will need to use the resources of several datacenters for the real implementation of AI instance in the current data processing paradigm. But, most probably, even several datacenters will have not enough resources for the implementation of AI instance. The power consumption of such implementation of AI instance would be enormously huge. It is possible to point out that current state of the art is unable to provide the basic hardware technologies for implementation of AI instance. Moreover, AI technologies need in another paradigm of data storing, representation and processing. The Nature implemented our brain as the unity of persistent storage and in-place data processing. It is possible to consider our brain model like a hint that show the necessity to merge the persistent memory and Processing Unit into the unity with the goal to store and to process data in-place. However, such paradigm (persistent memory + PU) contradicts with the existing paradigm of storing and processing the data. Moreover, the existing programming model and the code executing paradigm is unable to survive for the case of merging the persistent memory and processing circuitry into the unity on the chip's die. Modern paradigm of code execution distinguishes a persistent binary image from instance of the process is ready for execution. Any executable code is stored like a file inside the persistent storage. Generally speaking, it needs to transform the executable binary image into a process of OS that can be available by CPU for execution. Every process has virtual address space is created with the goal to load the user data from persistent storage into volatile memory for any transformation operations. User data is represented as raw binary stream that can be split on items of some granularities (1/2/4/8/16/32/64/128 bytes). Only items with such granularities can be processed by modern CPUs. As a result, it is hard to imagine a possible model of using the {NVM + PU}-unit for current paradigm of code execution. However, it is worth to point out that the {NVM + PU}-unit is able to create the new types of Processing Unit are able to solve the fundamentally new problems in the environment of completely different execution paradigm. CPU-centric model of computation has a huge number of drawbacks that originate from memory hierarchy and obsolete architecture of the computing core. We are continuously cultivating the model of computing core that is following to the model of Turing machine. Generally speaking, such model can be imagined like execution of instructions sequence. Every instruction defines the transformation command, granularity and placement of data. Such fundamentally sequential model of computing generates more and more intensive exchange traffic by instructions and data between main memory and CPU core. The classic model of centralized execution core (CPU-centric model) is the critical bottleneck many years already that destroys any efforts to improve the whole system performance significantly. But the performance of the system is not the final and key point of the whole problem because our goal not to achieve the maximum possible heat generation. The most critical point that CPU-centric model prevents us from analyzing the Big Data and to make the decision for reasonable time. It is possible to consider the distributed data processing under MapReduce model like possible solution of the declared problem. However, first of all, not every problem can be processed by MapReduce model. Secondly, CPU-centric and algorithm-oriented workflow of the Turing machine inevitably results in enormous network traffic and impossibility to achieve significant breakthrough in data processing. The algorithm-oriented model of data processing is not the answer for the Big Data and AI problems. Nowadays it is time to consider the data-centric model of data processing. This model should have the data in the basis of a paradigm instead of algorithms. The knowledge of nature and internal structure of data has to define the transformation activity with data. If the data is cornerstone of a new computing paradigm then the persistent memory is the basic building block of any processing unit. It means that not CPU or algorithm is the starting point but the persistent memory array with some data item(s). Generally speaking, it is possible to imagine that some capacity of NVM memory (or even elementary memory cell) would have the dedicated PU. If someone imagines an elementary DPU (NVM memory + PU) is able to store some data item persistently and to service the transformation of data in-place then the whole system can be imagine like the DPU array. The whole volume of data in the system can be distributed between DPUs of the array. This approach is fundamentally different from CPU-centric model because the knowledge of data structure is placed on hardware level. Also it is worth to point out that the DPU array creates the infrastructure where every data item has the dedicated processing unit that can be represented by elementary ALU or Neural Network. The DPU array architecture offers the rich infrastructure for data self-organization and deep implementation of AI primitives and techniques. Every DPU is able to have the knowledge of data nature. First of all, it means that DPUs will be able to interact with each other and to build the relations or any other structures by itself. Secondly, DPU is able to own the API for data item access or modification. This API can be received from the outside of DPU array or it can be synthesized by the DPU array. Nowadays, everybody expects algorithm-oriented model of data processing that looks like a sequence of instructions are initiated by end-user for execution on CPU core(s) is implementing the Turing machine instance. The data-centric environment creates the deeply decentralized model of data processing. This model doesn't imply the existence of any centralized initiator or manager of the data management workflow. The external world is represented by stimuli (for example, requested keywords or stored data) that manage the internal evolution of the DPU array by means of reaction and interaction of DPUs. Namely these external stimuli (in the form of saving a new data or requests with the goal to find some data, for example) define a current set of relevant goals that determine the evolution of the whole system (DPU array). The DPU array doesn't need in an algorithm in the form of the strict sequence of instructions like in the case of CPU-centric model. The data-centric model is based on the system evolution by means of active interaction and self-organization of DPU cores in the DPU array.

## III. RELATED WORKS

The data-centric processing is not completely new approach. The concept of mixing memory and logic has been around since 1960s. It is worth to mention such latest research efforts in near-data processing-in-memory area like Tesseract [1], HAMLeT [2], FlexRAM [25]. Junwhan Ahn et al. suggested a scalable processing-in-memory accelerator for parallel graph processing (Tesseract) [1]. Tesseract architecture represents the networks of cubes. Every cube is the HMC (Hybrid Memory Cube) that is composed of 32 vertical slices (called vaults). These vaults are connected by crossbar network. Each vault is composed of a 16-bank DRAM portion and a dedicated memory controller. A single core is placed at the logic die of each vault to perform computation inside DRAM. Host processors have their own main memory and Tesseract acts like an accelerator. Tesseract moves computation to the target core that contains data to be processed instead of allowing remote memory access. The movement of computation is implemented as a remote function call. Berkin Akin et al. suggested Hardware Accelerated Memory Layout Transformation (HAMLeT) system [2]. HAMLeT enables a software-transparent data reorganization capability performed in memory. Data reorganization operations appear as a critical building block in scientific computing applications such as a signal processing, molecular dynamic simulations, and linear algebra computations. The Hardware Accelerated Memory Layout Transformation (HAMLeT) system is integrated in the logic layer of a 3D-stacked DRAM similar to the Hybrid Memory Cube. The memory access stream is monitored to determine inefficient memory access performance by profiling the changes in the sequential DRAM addresses. The memory controller issues address remapping and data reorganization at runtime based on the monitored access pattern. HAMLeT's address remapping unit and data reshape unit handle address remapping and data reorganization completely in memory by avoiding data movement between the CPU and DRAM. Yi Kang et al. suggested the advanced intelligent memory system FlexRAM [25]. FlexRAM contains many very simple compute engines called P.Arrays that are finely interleaved with DRAM macrocells. To avoid incorporating extensive interconnection among P.Arrays, each P.Array sees only a portion of the on-chip memory. To increase the usability of P.Arrays, it is included a low-issue superscalar RISC core on chip. This processor, called P.Mem, coordinates the P.Arrays and executes serial tasks. Without the P.Mem, these tasks would need to be performed by the commodity microprocessor in the workstation or server (P.Host) at a much higher cost. Many FlexRAM chips can be connected to the commodity memory bus of a workstation or server. The high parallelism of the applications suggests including many P.Arrays. For this reason, it is supported only integer arithmetic. The neural network would benefit from multiplication support in P.Arrays. However, given the area cost of multipliers, it is best if several P.Arrays share a multiplier. Finally, to effectively support a wide range of applications, the P.Arrays work in a Single Program Multiple Data (SPMD) mode. The P.Host should start the P.Mems with a simple write to a special memory-mapped location. Next, the P.Host should pass the address of the routine to start executing in memory. And, finally, a master P.Mem should inform the P.Host when the job is completed. However, P.Mems cannot directly invoke the P.Host because memories cannot be masters of the memory bus. Consequently, to receive information from the P.Mems, the P.Host or the memory controller must poll on a location that the P.Mems can set. In-data processing-in-memory developed in parallel with near-data processing-in-memory research. Recently, emerging memory technologies such as resistive memory have become a focus. It is worth to mention such latest research efforts Memristive Boltzmann Machine [8], PRIME [10], Resistive TCAM system [18], MPIM [22], Pinatubo [28], ReGP-SIMD [31]. Mahdi Nazm Bojnordi et al. suggested the Hardware Accelerator for Combinatorial Optimization and Deep Learning (Memristive Boltzmann Machine) [8]. The Memristive boltzmann machine is an RRAM based hardware platform capable of accelerating combinatorial optimization and neural computation tasks. The key idea is to exploit the electrical properties of the storage cells and the interconnections among those cells to compute the dot product in situ within the memory arrays. The proposed accelerator resides on the memory bus and interfaces to general-purpose computer system via the DDRx interface. The Memristive boltzmann machine comprises a hierarchy of data arrays connected via a configurable interconnect network. A controller implements the interface between the accelerator and CPU. The data arrays are capable of storing the connection weights ($w_{ji}$) and the state variables ($x_i$). It is possible to compute the product of a weight and two state variables ($x_j x_i w_{ji}$) in situ within the data array. The interconnection network permits retrieving and summing these partial products to compute the energy change $\Delta E$ associated with a potential state update, and ultimately sends the $\Delta E$ results to the controller. Given the energy change that results from a potential update, the controller probabilistically decides whether to accept that update based on the Boltzmann distribution. Ping Chi et al. suggested the Novel Processing-in-memory Architecture for Neural Network Computation in ReRAM-based Main Memory (PRIME) [10]. Instead of adding logic to memory, PRIME utilizes the memory array themselves for computing. The add-on hardware in PRIME consists of simple modifications of the existing memory peripheral circuits with the goal to enable the computation function. PRIME aims to accelerate the Neural Network applications. It directly leverages ReRAM cells to perform computation without the need for extra PUs. PRIME partitions a ReRAM bank into three regions: memory subarrays, full function subarrays, and buffer subarrays. The memory subarrays only have data storage capability. The full function subarrays can execute Neural Network computations in computational mode. The buffer subarrays serve as data buffers for the full function subarrays. Qing Guo et al. suggested the resistive ternary content addressable memory systems for data-intensive computing (Resistive TCAM system) [18]. Resistive TCAM system is built on a DDR3-compatible DIMM. The system supports one or more TCAM DIMMs on the memory bus, each comprising an on-DIMM controller that serves as the interface to the DDR3 bus, translating conventional DRAM commands into TCAM operations. The processor communicates with the controller through a set of memory-mapped control registers (for configuring functionality), a key store (for buffering the search key), and a result store (for buffering the results). All of the chips share the on-DIMM command, address, and data buses; however, a search operation is restricted to operating on a single chip. Each chip has eight banks; a bank comprises a set of arrays that are searched against the search key, as well as a hierarchical reduction network for aggregating the results. Mapping an application to the TCAM system is a three-step process: (1) partition the workload between the processor and the TCAM system, (2) define the relevant data structures for storing the keys, and, (3) specify the TCAM operation to be performed on the keys. Mohsen Imani et al. suggested the multi-purpose in-memory processing using configurable resistive memory (MPIM) [22]. Multi-Purpose Resistive Memory (MPIM) consists of multiple memory banks which play a role of both memory and processing units. As a memory, MPIM can store and load general data. Stored data can be used for further processing. The main processor can request the data to be directly fetched from the persistent storage into MPIM instead of DRAM. MPIM includes several banks, where each bank contains crossbar memory which can be configured as either a memory mode for bitwise computation, or in CAM mode for search. MPIM can provide three key functionalities: (1) load/store operations as an additional memory next to DRAM, (2) nearest neighbor search operation, (3) bitwise computation (OR, AND, XOR). The target applications of MPIM is general streaming applications such as multimedia and graph processing. Shuangchen Li et al. suggested the processing-in-memory architecture for bulk bitwise operations in emerging non-volatile memories (Pinatubo) [28]. Bitwise operations are very important and widely used by database, graph processing, bio-informatics, and image processing. They are applied to replace expensive arithmetic operations. Pinatubo accelerates the bitwise operations inside the NVM-based main memory. Conventional computing-centric architecture fetches every bit-vector from the memory sequentially. The data walks through the narrow DDR bus and all the memory hierarchies, and finally is executed by the limited ALUs in the cores. Even worse, it then needs to write the result back to the memory, suffering from the data movements overhead again. Pinatubo performs the bit-vector operations inside the memory. Only commands and addresses are required on the DDR bus, while all the data remains inside the memory. To perform bitwise operations, Pinatubo activates two (or more) memory rows that store bit-vector simultaneously. The modified SA (Sense Amplifier) outputs the desired result. Thanks to in-memory calculation, the result does not need the memory bus anymore. It is then written to the destination address thought the WD (Write Driver) directly, bypassing all the I/O and bus. Amir Morad et al. suggested the resistive GP-SIMD processing-in-memory architecture [31]. ReGP-SIMD is a novel, hybrid general purpose SIMD computer architecture that addresses the issue of data synchronization (the data transfer between the sequential and parallel processors) by in-memory computing, through combining data storage and massively parallel processing. In resistive GP-SIMD, a novel resistive row and column addressable 4F2 crossbar is utilized. The architecture of the ReGP-SIMD processor includes a sequential CPU, a shared memory with two-dimensional access, instruction and data caches, a SIMD coprocessor, and a SIMD sequencer. The SIMD coprocessor contains a large number of fine-grain processing units, each comprising a single bit ALU (for performing bit-serial computations), single bit function generator, and quad, single bit register file. The GP-SIMD processor is thus a large memory with massively parallel processing capability. No data synchronization between the sequential and parallel segments is required since both the general purpose sequential processor and SIMD co-processor access the very same memory array. The execution time of a typical vector operation in GP-SIMD does not depend on the vector size, thus allowing efficient parallel processing of very large vectors. The sequential processor schedules and operates the SIMD processor via the sequencer. In a sense, the sequential processor is the master controlling a slave SIMD co-processor. Any common sequential processor may be used, be it a simple RISC or a complicated processor. At the very least, the selected processor should execute normal sequential programs. The SIMD coprocessor contains r fine-grain bit-serial Processing Units (PUs), where r is the number of rows in the shared memory array. Each PU contains a single bit Full Adder ("+"), single bit Function Generator ("Logic") and a 4-bit register file, RA, RB, RC and RD. A single PU is allocated per row of the shared memory array, and physically resides close to that row. The PUs are interconnected using an interconnection network. The set of all row registers of the same name constitute a register slice. Note that the length of the memory row (e.g., 256 bits) may be longer than the word-length of the sequential processor (e.g., 32 bits), so that each memory row may contain several words. Typical processing-in-storage architecture places a single or several processing cores inside the storage and allows data processing without transferring it to the host processor Intelligent SSD [5][13], Active flash [7], Active disk/iSSD [11], Minerva [12], BlueDBM [24], Smart SSD [26][34]. Leonid Yavits et al. suggested in-data processing-in-storage architecture (PRINS) [37]. This architecture progresses from random addressable to content addressable (associative) storage. PRINS enables massively parallel SIMD processing of the data inside the storage arrays. The processing is associative, making conventional in-storage processors redundant. There is no data transfer outside the storage arrays through a bandwidth limited internal SSD communication bus/network. The inherent performance (read/write access time and bandwidth) of the resistive memory can be utilized to the full extent, enabling very high computation throughput while reducing the energy consumption (mainly due to the lack of data movement inside storage). The main difference between PRINS and a hypothetical 3D near-data processing-in-storage architecture is the bitwise connectivity of memory and processing: In PRINS, each memory bit is directly connected to processing transistors, whereas in 3D near-data processing, the data must pass through memory interface circuits and through vertical interconnects, typically much fewer in numbers than the number of bits. In PRINS, the bulk of data ideally never leaves the memory. The computation is performed within the confines of the memory array.

## IV.  4. PiNVSM Architecture

Nowadays there are several methods of interaction with a computer system. These interactions can be classified into several fundamental ways. A user can provide input information by means of: (1) recording a sound; (2) recording an image/video; (3) use a tactile device; (4) use a pointing devices. Finally, trying to input the information into system means providing a sequence of symbols are recognizable by the system (Fig. 1). The goal of the system is to distinguish management requests from a user in the input. In other words, a management request is a service symbol that shows to the system what should be done with the following user data. The system can store, process or transform data according with a pre-established series of requisitions. Finally, a user expects to receive the data from the system by means of sound or video representation (Fig. 1). Also a user can expect some management activity inside of another system. Currently processor-oriented architecture is the dominating methodology. It means that processor-oriented system contains one or several CPUs (or CPU's cores), a memory hierarchy with main memory (DRAM) and non-volatile data storage (HDD, SSD and so on). At first, data should be extracted from the non-volatile storage into main memory and, then, CPU is able to process data by means of interchange data between small internal CPU's cache and main memory. Finally, data should be saved into the main memory after processing and, then, it should be flushed into the non-volatile storage. It is possible to say that such architecture is an algorithm-oriented architecture. Such architecture took place because of necessity to achieve compromise between existing technologies many years ago.

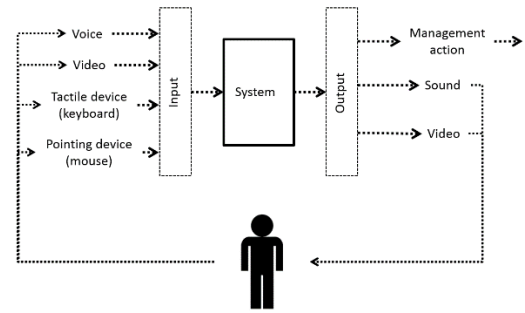

Fig. 1 Digital data processing ecosystem

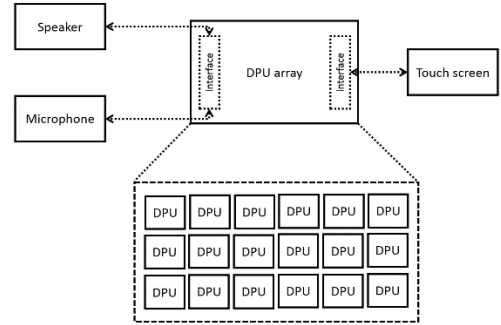

Fig. 2 DPU array like the basis of decentralized data processing architecture

Nowadays state of the art provides opportunities to build CPU with dozens of (and hundreds of) cores. Algorithm-oriented architecture becomes really inefficient for such technological environment. Dedicated or Application Specific Integrated Circuits (ASIC) seems to be the ideal development platform to explore different architectures for the emerging NVM. Future computational systems should be data-oriented systems. It means that: data shouldn't be transferred between storage and operation places, instead operation core should be in the storage place, or storage devices. It is possible to imagine an elementary Data Processing Unit (DPU) that can contain processing unit and some amount of NVM memory. One practical and well-known storage system is the so-called Processor-In-Memory (PIM) devised by Patterson from Berkeley, and currently demonstrate as its iRAM project [40]. Finally, the whole system can be imagined as DPU arrays (Fig. 2). The goal of DPU is to be an active basic element of the system, that can store data, providing access and been capable of manipulating some piece of data in the system. In other words, DPU can be treated as storage space with some capacity of processing and/or with dedicated embedded PU devoted to data processing in the storage device. The goal of DPU array is to provide interaction space between DPUs for interaction during data processing.

Every DPU is the combination of NVM memory with dedicated Processing Unit (Fig. 3). Such NVM memory array can be accessed for data processing by means of requests to the dedicated Processing Unit. The goal of synthesis is the Processing Unit and some capacity of NVM memory is to make possible data transformation in the place of storing (offload operations with data). The concept of active core (PU + NVM memory) creates the environment for decentralized data processing paradigm when the DPU array will be able to operate with data completely independently in every DPU. The goal of active core architecture is to change the concept of passive memory (keeping data only) by the paradigm of active DPU array is able to transform the data and to create the relations between the data in different DPUs. The core (PU) dedicated to the piece of data (NVM memory) is able to exclude the necessity to use the synchronization primitives because all transformation requests can be placed into the one queue that will be executed by one dedicated core (PU).

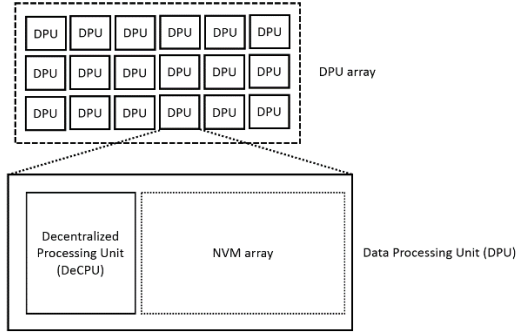

Fig. 3 DPU is the combination of NVM memory and processing unit

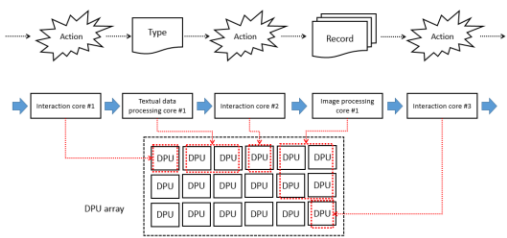

Fig. 4 Interaction sequence workflow

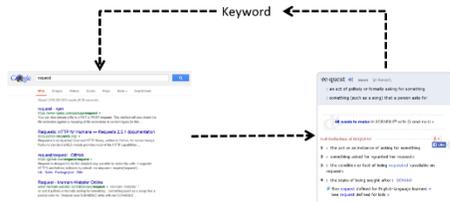

Fig. 5 Keyword like the new basis of data processing paradigm

A user interacts with the computer system by means of management actions. Such management action can be touching sensor screen, clicking an object's icon and so on. The management action is a service symbol that requests some system's activity by means of storing some data, transformation data or providing access to data. As a result, interaction of a user with the system can be imagined as sequence that combines as management actions as manipulation by data (Fig. 4). DPUs in array can be appointed for processing requests in the interaction sequence (Fig. 4).

Usually, a keyword is a starting point for operation with information. A keyword can be used for searching information, for getting access to information, for storing and transformation of information. It is possible to say that keyword-based way of interaction with search engine is de-facto the user model of operations with information (Fig. 5).

What is text?: Text is a set of identifiable keys.

What is interface with a system?: Interface is a set of identifiable keys.
A. Keyboard provides identifiable symbols (as service as data symbols).
B. Mouse provides position (or location) of identifiable object.
C. Voice is oral representation of text.
D. Video is a set of recognizable or identifiable images (objects).

## Text == System Interface

Fig. 6 Text is the fundamental basis of data representation and interaction with user

But the keyword is a part of any textual information representation. Moreover, textual representation of the information is the most natural information transmission and organization method used by mankind. Finally, text can be a fundamental user interface in decentralized data processing architecture (Fig. 6).

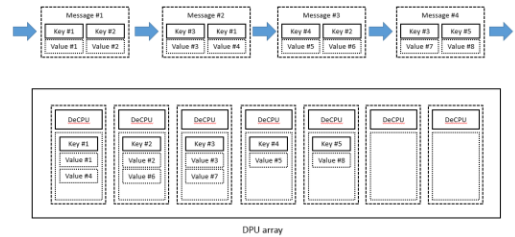

Fig. 7 Data-oriented processing

Textual representation of information can be treated as aggregation (or ordered structure) of keyword-values set. It means that it is possible to extract keywords from any text. The rest of the text is values are associated with keywords. The system can receive textual messages in sequential or concurrent manner (Fig. 7). Data-oriented processing means the splitting a message on keyword-value pairs and to store every keyword-value inside of NVM memory of dedicated DPU.

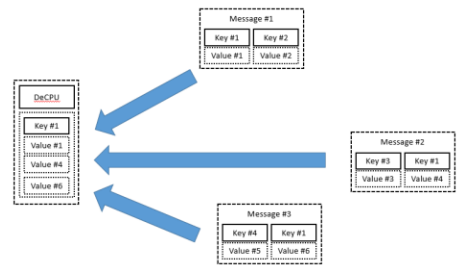

Fig. 8 Data self-organization

DPU stores and processes information related to a concrete keyword. It means that NVM memory of a concrete DPU will gather values for the same keyword of different messages (Fig. 8). Such technique can be considered like the basis for data self-organization.

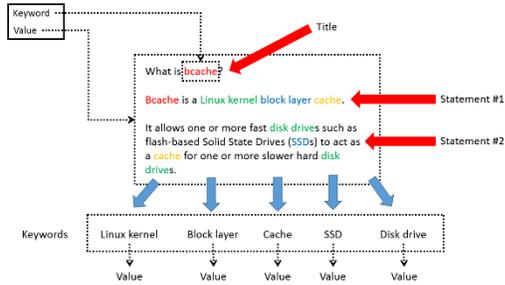

Fig. 9 Keyword-Value basis of analysis and synthesis of textual data

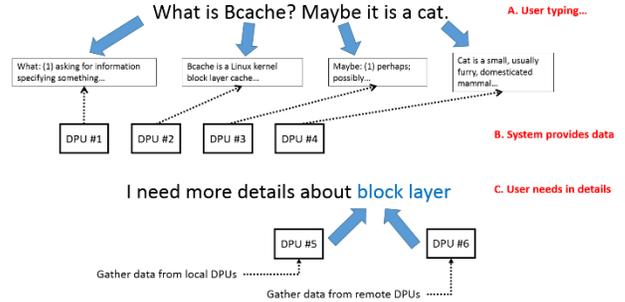

Fig. 11 Model of user interaction with decentralized DPU array

Textual data is the aggregation of keywords (Fig. 9). It means that any text can be decomposed on the set of keywords are combined in the unity by means of relations. Any data is the combination of items are united by means of relations of items. If a user saves some data in the system it means that the data can be decomposed on keyword-value pairs by means of analysis. The analysis provides the opportunity to decompose the data on items that can be stored in the DPU array in the most natural and efficient manner (keyword-value pair). The storing and representation of data in the form of keyword-value pairs creates the rich opportunities for synthesis of data. The keyword-value pair is the efficient basis for data self-organization and the creation of a new knowledge by DPU array itself without any special instructions or algorithms.

DPU array will be efficient environment for data interaction, organization and processing. It will provide natural support of model of user interaction with information.

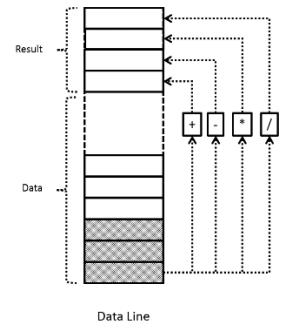

Fig. 12 Data line concept

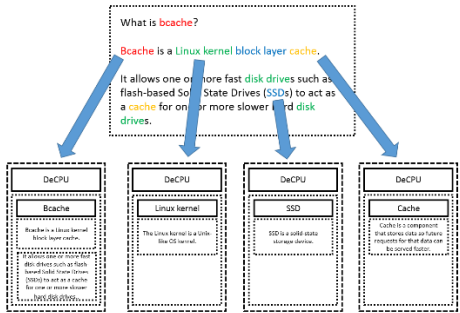

Fig. 10 Decentralized DPU array like the natural way of textual-based data processing

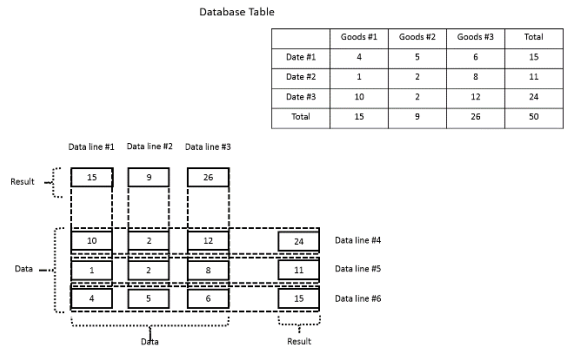

Fig. 13 Example of using data lines for database table

Text-like structure represents the aggregation of keyword-value pairs are joined into the unity by means of relations. If one keyword-value pair is stored into an independent DPU then the DPU array represents the media of storing data in the form of keyword-value pairs. The active nature of DPU's core provides the opportunity to create relations between keyword-value pairs in the decentralized manner with the goal to store the data or to create the new knowledge. Machine learning approaches could be a basis for keyword/value pairs extraction. A value is associated with keyword contains dependent keywords that can be a starting point for information gathering and storing. If a user is working with some text then dependent keywords are potential user's information requests.

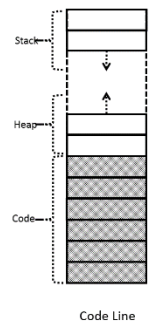

Fig. 14 Code line concept

NVM space could be shared between data and code lines. Data line is an array of data items with associated operations (Fig. 12 - Fig. 13). Code could be simply stored as data and it can be treated as data line. Otherwise, it should be copied into special place that will be treated as code line. Code line is a sequence of operations with the prepared environment that can be used for transformation of data lines (Fig. 14). Finally, after execution, code line can be transformed in data line that will store results of data transformation.

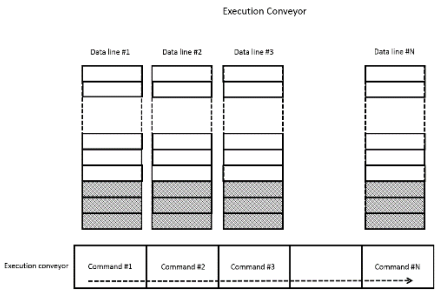

Fig. 15 Execution conveyor in NVM memory

Sequence of commands in a code line could be imagined as execution conveyor (Fig. 15). Every command describes one elementary operation in sequential execution flow. And every command could be applied to different data lines in concurrent manner.

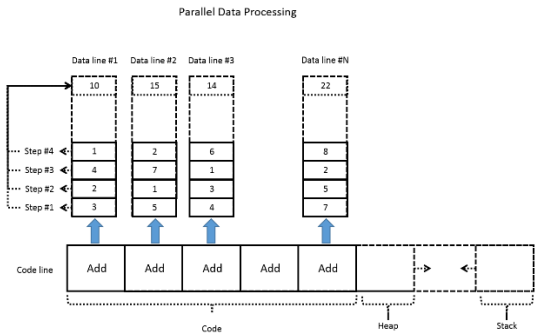

Fig. 16 Parallel data processing in data lines

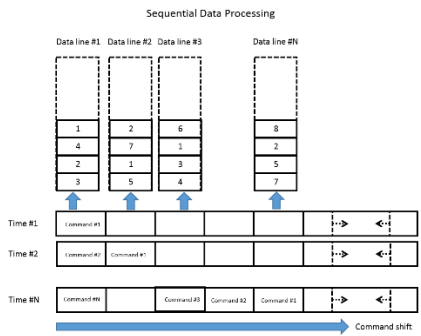

Fig. 17 Sequential data processing on the basis of execution conveyor in NVM memory

Let's imagine that it needs to calculate the total numbers for all columns in a table. It means that it needs to apply the calculation operation several times from the first row to the last one in the every column. The sequence of commands should contain the addition operation for every column. And this operation should be repeated before achieving the result of the whole code execution (Fig. 16).

It is possible to use command shift approach for the case of necessity to execute more complex operation with the data in a row (Fig. 17). And different steps of the same transformation operation can be made concurrently in several data lines.

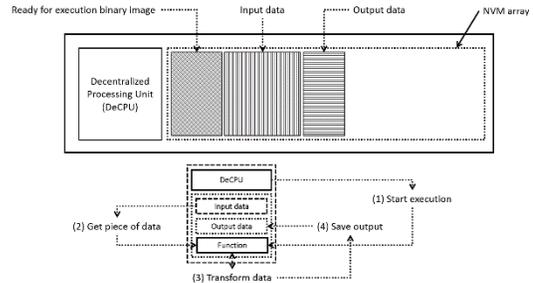

Fig. 18 Concept of data processing in the DPU

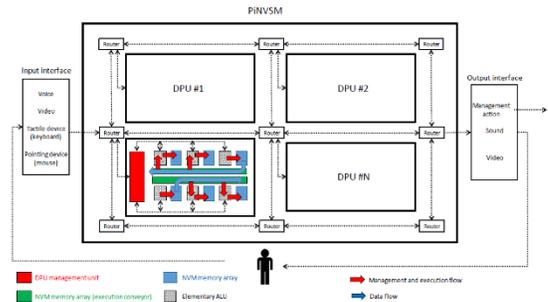

Fig. 19 DPU architecture

Keyword is the basis for data organization, relation creation, and searching the data. However, the value is associated with a keyword can be imagined like binary stream that can be stored in one or several data lines. Generally speaking, DPU could distinguish some structure in the value (for example, value could be organized like the sequence of keywords). As a result, items of the value can be stored in the different data lines are organized in some way. The executable code could be represented in the form of keyword-value pairs likewise ordinary data. For example, single method can be represented by keyword-value pair where the name of method is the keyword and the value is the binary code. The method (or binary code) can be represented in the form of data lines during regular storing. First of all, DPU array is the storage space where every DPU can be imagined like the bucket is storing one or several keyword-value pairs. The keywords represent the basis for data organization, searching and transformation in the DPU array. However, any transformation needs in code that is able to transform the data. It means that such code needs to be found in the DPU array by means of keyword(s). For example, the name of method can be used for searching a necessary code. If the position of data lines with the code was found in some DPU then the content of data lines needs to be copied into the DPU that keeps the data is awaiting the transformation. Also the found code could be transformed into the code line(s) with the goal to prepare the code for the real execution. NVM memory can be used as for keeping data lines as for execution conveyor that is able to receive the code line and to apply the sequence of instruction by elementary ALU on data line (Fig. 19).

## V. DISCUSSION

The AI problem has no solution in the environment of existing hardware stack and OS architecture. First of all, the power consumption of such implementation of AI instance would be enormously huge. Also the fundamentally sequential model of computing (model of Turing machine) generates more and more intensive exchange traffic by instructions and data between main memory and CPU core. Moreover, it is worth to mention that CPU-centric model of computation has a huge number of drawbacks that originate from memory hierarchy and obsolete architecture of the computing core. The CPU core transforms data by means of small set of really fast registers. And all modern programming languages are based on the programming model is taking into account the limited set of CPU core's registers. As a result, the key assembler instruction is the exchanging by data between the main memory and registers of CPU core. The important issue of CPU-centric computing is the L1/L2/L3 caches that create the problem of coherence of data between the main memory and the caches of different CPU cores. The modern CPU spends a lot of time for data/code exchange between the main memory and CPU cores. The tricky problem of modern OS is that any file system provides very important features (namespace, permissions) via the inevitable necessity to copy and convert metadata between persistent storage and internal OS's metadata representation in the volatile memory. This principal dichotomy between volatile and persistent memory has fundamental nature in the modern OS and it cannot be resolved by simple addition of a new OS's subsystem. The simple exchange the volatile memory (DRAM) on non-volatile memory cannot resolve the problem because the internal mechanisms of OS and file system require the presence of two levels of memory (as minimum) and exchanging by data between of them. Finally, would the PiNVSM architecture resolve all above-mentioned issues of CPU-centric computing paradigm?

The basis of PiNVSM architecture is the concept of DPU that contains the NVM memory and dedicated PU (elementary ALU, Neural Network engine and so on). Generally speaking, all necessary PU's registers can be implemented in the space of NVM memory. It means that power-consumption can be decreased dramatically in the case of using the NVM memory like ALU's registers because power-hungry DRAM/SRAM will be not used for keeping the data during transformation. Moreover, PiNVSM architecture doesn't need to transfer data between the place of storing and place of data processing because NVM memory of DPU is the single space for storing and transformation of data. Generally speaking, it means that DPU is the basis for exclusion the significant amount of moving instructions. Finally, as a result, the exclusion of moving instructions is able to decrease the power-consumption because it provides the opportunity to process the data in-place.

PiNVSM architecture creates completely unique paradigm is able to overcome the limitations as Turing machine model as von Neumann architecture. First of all, PiNVSM architecture is based on the DPU array. Every DPU is the independent unit that represents as persistent space for data as active processing unit is able to process data in this persistent space. It means that DPU array hasn't a centralized computing core that needs to receive the code/data for any operations. Every data portion has dedicated computing core in PiNVSM paradigm that excludes the necessity to transfer data to the place of data processing. Does PiNVSM architecture need to transfer the code to the place of data processing? If it is possible to detect the nature of data through the storing process then it is possible to prepare the related code near the data location. Otherwise, if the data nature is undetectable beforehand or the code doesn't exist in the DPU array yet then it is possible that a code needs to be transferred to the data location at the moment of data transformation request. However, such operation of code transfer would take place inside of the DPU array locally. Finally, if the code has been transferred from one DPU into another one then it doesn't need to repeat such transfer operation again. The destination DPU is able to store the code for the future requests. Moreover, it is possible to imagine that frequently requested code can be converted into hardware circuitry by means of FPGA programming, for example.

The combination of data portion with dedicated computing core in the DPU creates the way to avoid the Turing machine's limitations. The key problem of Turing machine is the strict sequential flow of instructions execution because the whole approach is algorithm oriented. The foundation of Turing machine is the executor that is able to know what it needs to do after the receiving the next instruction code. The instruction code defines the operation, the granularity of data, and the data location. PiNVSM paradigm follows the completely different flow. The basis of PiNVSM paradigm is not algorithm but data itself. Every DPU contains data portion that is associated with the set of keywords. Generally speaking, DPU array creates deeply decentralized paradigm of data processing. The first step of data processing in DPU array is the definition of keywords that select the DPUs for applying a requested operation. Moreover, the knowledge of keywords by DPUs can be a basis for code synthesis or internally initiated data transformation in the DPU array. As a result, data processing takes place in every selected DPUs in decentralized manner without the necessity to follow the sequential flow of instructions like in the Turing machine. However, the Turing machine model can be used inside the DPU by means of applying an algorithm for one data portion by dedicated computing core. But if the computing core is represented by a Neural Network engine then the Turing machine model can be completely excluded from using in the PiNVSM paradigm. The important point of DPU array that every portion of data has own dedicated PU. Finally, it means that independent portion of data in different DPUs can be processed concurrently (physically at the same time) without the necessity to follow the sequential execution model of the Turing machine. However, if some data portion in the DPU needs to be processed by means of several sequential steps then the execution flow needs to be sequential.

The PiNVSM paradigm excludes the necessity to use the memory hierarchy because the DPU concept unites the NVM memory (persistent space) and Processing Unit (ALU or Neural Network) into the one unit. It means that data is stored persistently in the place where the data transformation would take place by means of dedicated core. The critical issue of memory hierarchy is the necessity to copy data between different levels with the goal to access or to transform the data. The PiNVSM architecture is free from such drawback because the PU has direct access to persistent representation of data in byte-addressable manner. Moreover, any complex data structure can be split on elementary items that can be stored into independent DPU with dedicated computing core(s). Finally, it means that one DPU is able to apply the elementary transformation on one item. But the DPU array is able to make the transformation of complex structure by means of concurrent execution of elementary transformations in different DPUs.

The PiNVSM architecture suggests a principally new architecture of the computing core that creates a new opportunity for data self-organization, data and code synthesis. First of all, every DPU has dedicated NVM memory array. This dedicated NVM memory is the basis for different type of DPU's internal structure. Particularly, the persistent space (NVM memory) can be used by ALU (or any other type of PU) like any number of registers with any possible granularity. As a result, it means that PiNVSM architecture resolves the limitation of obsolete model of modern CPU's registers. Moreover, the using of NVM memory is able to decrease the power-consumption of the processing core(s). The using of DPU's dedicated NVM memory for placing the registers of PU opens principally new opportunities. Because such model of PU's registers is able to completely exclude the necessity to copy data from the place of storing to the place of processing. The placement of some data portion in DPU's NVM memory can be represented like a register with initial state of data. Also some free space of NVM memory can be represented like a destination register for the data transformation operation. Generally speaking, DPU's internal NVM memory is able to mimic the obsolete registers' model of modern CPU because the most of programming languages expect this limited set of registers. However, DPU architecture provides very flexible way for creation of any registers' model.

The DPU architecture completely excludes the coherence problem between the main memory and L1/L2/L3 caches. It needs to take into account that the persistent space (NVM memory) of DPU is used like single space for storing data and allocation of registers. Generally speaking, the DPU doesn't need in any level of caches because the place of storing data and place of data processing is coinciding in PiNVSM paradigm However, if the DPU could contain several cores (ALU or Neural Network) then it needs to have some technique of synchronization the access different cores to the same piece of data. This protection can be achieved by using the only one core in the DPU or by means of safe allocation of cores' registers in the NVM memory of DPU.

File systems are very critical service in the modern world. Namely, file systems provide access to our data. However, classical file system's model is already obsolete and is unable to satisfy the needs of modern community. Nowadays, namespace concept of the file system is very critical for us like human beings because searching information on the keywords basis is the culture of everyday life. However, the concept of hierarchy of files provides a lot of limitations that make practical impossible the easy and flexible way of data reorganization. Also the file concept plays important role in the performance degradation of modern systems.

Usually, file system provides a namespace in the form of files' hierarchy that can be found and accessed on the name basis. The PiNVSM architecture creates fundamentally different namespace model in that every data portion (is kept in DPU) has associated set of keywords. These keywords can be used for creation of relations of any complexity between data portions. The DPU array approach creates very flexible model of namespace that provides opportunity to build the namespace of any complexity on the basis of representative set of keywords. Generally speaking, PiNVSM architecture rethinks the model of file system's namespace by means of providing the flexible way of building namespace on the basis of grouping the data portions that satisfy to the representative set of keywords.

## VI. CONCLUSION

The AI problem has no solution in the environment of existing hardware stack and OS architecture. CPU-centric model of computation has a huge number of drawbacks that originate from memory hierarchy and obsolete architecture of the computing core. Also the fundamentally sequential model of computing (model of Turing machine) generates more and more intensive exchange traffic by instructions and data between main memory and CPU core. Additionally, the power consumption of CPU-centric implementation of AI instance would be enormously huge.

This paper suggests the concept of Processor in Non-Volatile Memory (PiNVSM) architecture. The basis of PiNVSM architecture is the concept of DPU that contains the NVM memory and dedicated PU. All necessary PU's registers can be implemented in the space of NVM memory that can decrease the power-consumption dramatically. NVM memory of DPU is the single space for storing and transformation of data. It means that DPU is the basis for exclusion of significant amount of moving instructions. The DPU architecture completely excludes the coherence problem because the place of storing data and place of data processing is coinciding in PiNVSM paradigm. In the basis of PiNVSM architecture lies the DPU array is able to overcome the limitations as Turing machine model as von Neumann architecture. The DPU array hasn't a centralized computing core. Every data portion has dedicated computing core that excludes the necessity to transfer data to the place of data processing. The basis of PiNVSM paradigm is not algorithm but data itself. Every DPU contains data portion that is associated with the set of keywords. Generally speaking, DPU array creates deeply decentralized paradigm of data processing. Any complex data structure can be split on elementary items that can be stored into independent DPU with dedicated computing core(s). Finally, it means that one DPU is able to apply the elementary transformation on one item. But the DPU array is able to make the transformation of complex structure by means of concurrent execution of elementary transformations in different DPUs. The PiNVSM architecture suggests a principally new architecture of the computing core that creates a new opportunity for data self-organization, data and code synthesis.